\documentclass{llncs}
\normalsize

\usepackage[ruled]{algorithm}
\usepackage{algorithmic}
\usepackage{textpos}
\usepackage{cite}

\usepackage{makeidx}  
\usepackage{latexsym}
\usepackage{graphicx}
\usepackage{graphics}
\usepackage{longtable}
\usepackage{multirow}
\usepackage{verbatim}
\usepackage{listings}
\usepackage{threeparttable}
\usepackage{subfig}
\usepackage{xspace}
\usepackage[latin1]{inputenc}
\usepackage{epsfig}
\usepackage{pifont}
\usepackage{url}
\usepackage{rotating}
\usepackage{color}

\newcommand{\fig}[1]{Figure~\ref{#1}}

\newcommand{\tbl}[1]{Table~\ref{#1}}

\begin{document}



\title{An Experience Report of Large Scale Federations}

\author{Andreas Schwarte\inst{1} \and Peter Haase\inst{1} \and Michael Schmidt\inst{1} \and \\ Katja Hose\inst{2} \and Ralf Schenkel\inst{2}}
\tocauthor{Peter Haase, Michael Schmidt, Andreas Schwarte, Katja Hose, Ralf Schenkel}

\institute{fluid Operations AG,\\
69190 Walldorf, Germany,\\
\email{firstname.lastname@fluidops.com}
\and
Max-Planck-Institut f\"{u}r Informatik\\
66123 Saarbr\"{u}cken, Germany,\\
\email{khose@mpi-inf.mpg.de}, \email{schenkel@mpi-inf.mpg.de}}

\maketitle

\begin{abstract}
We present an experimental study of large-scale RDF federations on top of the Bio2RDF data sources, 
involving 29 data sets with more than four billion RDF triples deployed in a local federation. 
Our federation is driven by FedX, a highly optimized federation mediator for Linked Data. 
We discuss design decisions, technical aspects, and experiences made in setting up and optimizing
the Bio2RDF federation, and present an exhaustive experimental evaluation of the federation scenario.
In addition to a controlled setting with local federation members, we study implications arising in a hybrid
setting, where local federation members interact with remote federation members exhibiting higher
network latency. The outcome demonstrates the feasibility of federated semantic data management
in general and indicates remaining bottlenecks and research opportunities that shall serve as a
guideline for future work in the area of federated semantic data processing.
\end{abstract}

\section{Introduction}
\label{sec:introduction}

The vision of the Semantic Web, i.e.~transforming the current Web of Documents into a Web of Data, has been gaining more and more attention lately. Connecting not only documents on the web but establishing connections on the data level, opens up new possibilities of automatic interaction, knowledge representation, question answering, and knowledge acquisition that has not been available before. Especially, the Linked Open Data~\cite{BizerHB09} community has been working on providing links between RDF data on the Web -- making RDF and SPARQL the popular standards for data representation and querying on the Semantic Web. 

The Linked Open Data cloud 
now consists of 295 data sources and about 31 billion 
RDF triples -- and is constantly growing. One of the core principles of Linked Data is to use Uniform Resource Identifiers (URIs) as unique identifiers that globally represent a specific entity and can be used across data sources to interlink resources. As the data provided on the Web and by each source is rapidly outgrowing the capacity of purely explorative querying --- DBpedia 
for instance now has about 1 billion triples --- some sources provide their data collections for download as RDF dumps or enable access via SPARQL endpoints. 
Accessing a data set through its SPARQL endpoint has two major advantages over downloading RDF dumps. First, it allows to evaluate complex queries over the data set without the need to set up a private triple store, possibly even on expensive high-end hardware. Second, data behind SPARQL endpoints is often more up-to-date compared to available dumps (which may be updated only in large intervals and therefore not include recent updates to the data set). 

For queries that include multiple data sets, connecting multiple SPARQL endpoints to a federation comes with a number of benefits over a centralized integration in a single triple store: (i) the data is always up-to-date; (ii) the computational load is shared among servers (which holds even for local federations, where different data sets are kept in different triple stores on local servers); (iii) available endpoints can be integrated into federations ad hoc, avoiding the often time-consuming process of loading dumps into local repositories; and (iv) increased flexibility, allowing to use and query arbitrary combinations of the data sources in different, requirement-tailored federations. The latter two are particularly important when local
 data sources are combined with public endpoints.  

As indicated by our experimental results in this paper, for typical queries against large
federations effectively only a small subset of the endpoints contribute to the final result.
Consequently, splitting up a query into subqueries and evaluating them in parallel over a local federation can be
even faster than evaluating the full query over a single triple store containing all the data of all sources
(compare, for instance, the experimental results in~\cite{schmidt2011a}).
There remains, of course, a tradeoff between the benefit of distributed and parallel processing and the communication overhead between different instances, so that some queries would be evaluated more efficiently in a centralized setup.

Having seen many promising results in previous benchmarks on federated query processing with
only few federation members and in the order of a hundred million triples~\cite{schwarte2011a,gorlitz2011splendid,schmidt2011a},
the goal of this paper is to demonstrate the practicability of large-scale RDF federations: using FedX~\cite{schwarte2011a},
a highly optimized Linked Data federation mediator, we set up a federation with $29$ SPARQL endpoints
hosting the individual data sets from the Bio2RDF domain, containing more than $4$ billion RDF triples in total.
In our experimental results, we study the performance of queries against such federations, comparing
 (i)~local federations with all SPARQL endpoints running on servers in a local network and (ii) hybrid federations where some sources are hosted locally and others on the
Web or on Amazon EC2; 
the latter setup is a classical setting in
the enterprise context, where companies need to combine local, private data sources with open data
accessible through public SPARQL endpoints. It is beneficial whenever local working copies of some data
(e.g., generated based on information extraction from natural language text, latest experiments,
user-generated/corrected data, downloaded cleaned dumps, etc.) need to be augmented with public
information.

{\bf Contributions.} In summary, we make the following contributions.
\begin{itemize}
\item We present the first real large-scale federation setup in the context of RDF, implemented using the FedX federation mediator on top of $29$ Bio2RDF SPARQL endpoints containing about~$4.1$ billion RDF triples. 
\item Our description summarizes problems and solutions, as well as practical aspects of the federation setup. All experiments can be reproduced by anyone following the instructions outlined in Section~\ref{sec:experiments}.
\item We set up a public demonstrator of our Bio2RDF federation, supporting live queries against and browsing of
the underlying federated data graph.\footnote{See \url{http://biofed.fluidops.net}}
\item An exhaustive evaluation along different dimensions -- including technical setup aspects, performance and scalability, and network latency -- proves the feasability of federated RDF data management in large-scale settings.
\item Our experiments reveal open issues and current limitations, which serve as a guideline for future work in the area of federated semantic data processing.
\end{itemize}
{\bf Structure.} After a discussion of related work in Section~\ref{subsec:relatedwork}, we turn towards a description of the federation technology, the FedX system, in Section~\ref{sec:technology}. Next, Section~\ref{sec:federation-setup} describes the federation setup, including a motivation of the chosen scenario, a description of the datasets, and a general discussion of the benchmark queries. In Section~\ref{sec:experiments} we describe the infrastructure setup, motivate the different experimental scenarios, metrics, and present an exhaustive discussion of the experimental results. Finally, we elaborate on the implications of our results for future work and conclude with some final remarks in Section~\ref{sec:conclusions}.

\subsection{Related Work}
\label{subsec:relatedwork}

With the uptake of Linked Data in recent years, the topic of integrated querying over multiple distributed data sources has attracted significant attention. In order to join information provided by these different sources, efficient query processing strategies
are required, the major challenge lying in the natural distribution of the data.
So far,  the commonly used approach for query processing in large scale integration scenarios  is still to integrate relevant data sets into a local, centralized triple store.
Examples of such integrated repositories are the LOD cloud cache\footnote{\url{http://lod.openlinksw.com/}} or Factforge\footnote{\url{http://factforge.net/}} that integrate significant subsets of the Linked Open Data cloud. As a more domain specific example,  Linked Life Data\footnote{\url{http://linkedlifedata.com/}} integrates 23 datasources from the biomedical domain.
Following a similar approach, the OpenPHACTS project\footnote{\url{http://www.openphacts.org/}} attempts to build an integrated resource of multiple databases in the pharmaceutical space.

Yet recently one can observe a paradigm shift towards federated approaches over the distributed data sources
with the ultimate goal of virtual integration \cite{Hartig09ExecutingSparqlQueries,ladwig2011} . A recent overview and analysis of federated data management and
query optimization techniques is presented in \cite{goerlitz2011}. 

Basic federation capabilities have been added to SPARQL with the SPARQL 1.1 Federation extensions\footnote{\url{http://www.w3.org/TR/sparql11-federated-query/}}. They introduce the SERVICE operator, which allows for providing source information directly within the SPARQL query. 
Aranda et al.~\cite{aranda2011} provide a formal semantics for the language extensions. 
While our federation approach in FedX also supports SPARQL 1.1 Federation, it does not require these extensions. Instead, it is fully compatible with the SPARQL 1.0 query language, i.e. multiple distributed data sources can be queried transparently as if the data resided in a virtually integrated RDF graph. Source selection is achieved through automated means over a set of defined sources (which can be dynamically extended) without explicit specification in the query.

In \cite{schmidt2011a} we introduced the FedBench benchmark suite for testing and analyzing the performance of federated query processing strategies. Our experiments presented in this paper build upon the FedBench benchmark, but evaluate a federation scenario of a significantly larger scale.

\section{FedX}
\label{sec:technology}

In the following we give some insights into the technologies and concepts of FedX~\cite{schwarte2011a}, which is used
as the federation mediator in our experimental study. FedX is a practical framework for transparent access to Linked Data sources through a federation. By virtually integrating multiple heterogeneous sources, the federation mediator exposes the union of all source graphs transparently to the user, i.e. the user can evaluate queries as if the data resided in a single triple store. Federation members are specified as a list of SPARQL endpoints, which can be added to (or removed from) the federation on-demand, since no precomputed statistics are required for query processing. With its federation-tailored optimization techniques discussed below, FedX enables an efficient and scalable SPARQL query processing for different practical federated settings.

\begin{figure}[b]
\begin{center}
\includegraphics[width=0.9\textwidth]{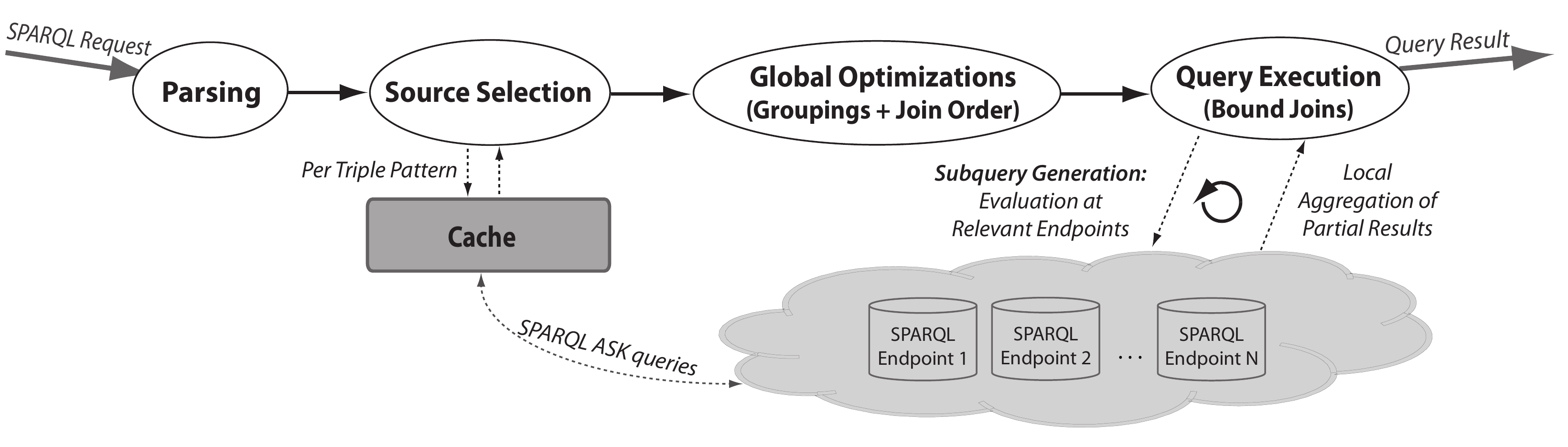}
\end{center}
\vspace{-0.4cm}
\caption{Federated Query Processing Model of FedX }
\label{fig:fedx_model}
\end{figure}

The query processing workflow in FedX is depicted in \fig{fig:fedx_model}. FedX first parses the query into an internal tree-like representation, which is then optimized using various techniques. Optimization in FedX includes \textit{source selection} (i.e., finding the relevant sources for each triple pattern using SPARQL ASK requests), forming \textit{exclusive groups} (i.e., grouping those triple patterns that have the same single source), and a rule-based \textit{join reordering} approach. At runtime FedX manages a so-called \textit{source selection cache}, containing information about which endpoints can potentially yield results for a given triple pattern. With this cache, FedX is able to reduce the number of requests since it can prune endpoints that are not relevant for the evaluation of subqueries directly.


%
%

As a user-facing frontent built on top of the federation managed by FedX, we provide a browser-based demo system based on the Information Workbench\footnote{http://www.fluidops.com/information-workbench/}, a Linked Data platform which allows to declaratively use widgets within a semantic wiki to interact with the underlying Linked Data graph. 

\vspace*{-0.25cm}
\section{Experiment Scenario: Federating Bio2RDF}
\label{sec:federation-setup}
\vspace*{-0.25cm}
For our experimental study of a large scale federation we decided to use data sets from the life science domain. The industries in the life sciences (including pharmaceuticals, bio technology) have been an early adopter of semantic technologies and the value of providing integrated access to distributed data sources has been demonstrated in many practical applications \cite{DBLP:journals/semweb/Dumontier10}. Most of the Linked Data data sets in the life sciences  have been published as part of the Bio2RDF initiative, with the goal to provide interlinked life science data to support biological knowledge discovery. Compared to other domains, the data sets that have been developed for the life science domain are of rather high quality and very well interconnected. Consider as an example the Drugbank dataset which provides direct links for most drugs to the corresponding KEGG compounds.

For our federation we have selected 29 data sets, covering -- to the best of our knowledge -- all relevant publicly available data sets in the domain. In total, the selection comprises more than 4 billion triple.  Table \ref{tbl:datasets} lists all datasets and depicts the number of triples and entities, as well as the main instance type(s). 

\vspace*{-0.45cm}
\begin{table}[h!]
\scriptsize
\begin{tabular}{|l|l|l|l|l|l|l}
    \hline
 \# &   \textbf{Dataset} 	& 	\textbf{\#Triples}	& \textbf{\#Entities}	& \textbf{Instance type(s)} \\ \hline
    
1 &		CellMap				& 	149k			&	60k		& biopax-2:protein \\ \hline
2 &		ChEBI 				&  	650k		&	238k	& - \\ \hline
3 &		DailyMed 			&  	163k		&	68k		& dailymed:drugs \\ \hline
4 &		Disease Ontology	& 	145k		&	110k	& - \\ \hline
5 &		DBpedia Subset 		&	70M			&	31M		& e.g. dbo:Drug \\ \hline
6 &		Diseasome 			& 	75k			&	30k		& diseasome:genes \\ \hline
7 &	    DrugBank 			& 	0.5M		&	290k	& drugbank:drugs \\ \hline
8 &		Entrez-Gene			&	161.5M		&	67M		& entrezgene:Gene	\\ \hline
9 &		Genewiki 			&	1.0M		&	391k	& - \\ \hline
10&		KEGG 				& 	2.4M		& 	1M		& kegg:Compound, kegg:Drug, kegg:Enzyme, kegg:Reaction \\ \hline
11&		Mappings 			&	2.8M		& 	4.1M	& - \\ \hline
12&		Pubmed 				& 	1.4B		&	299M	& pubmed:Citation	\\ \hline
13&		UMLS 				&   121M		&	27.7M	& skos:Concept	\\ \hline
14&		Uniprot				& 	2.3B		&	495M	& uniprot:Protein, uniprot:Journal \\ \hline
15&		BiogGRID 			&	12M			&	4.7M	& biopax-2:protein \\ \hline
16&		Gene Ontology 		&	320k		& 	187k	& skos:Concept \\ \hline	
17&		HapMap 				&	22M			&	43M		& - \\ \hline
18&		HPRD 				&	2M			&	777k	& biopax-2:protein \\ \hline
19&		Humancyc 			&	327k		&	143k	& - \\ \hline
20&		IMID 				&	83k			&	36k		& biopax-2:protein\\ \hline
21&		IntAct 				&	16.6M		&	5.5M	& biopax-2:protein \\ \hline
22&		LHGDN 				& 	316k		&	160k	& - \\ \hline
23&		LinkedCT 			& 	7.0M		& 	2.8M	& linkedct:trials, linkedct:condition \\ \hline	
24&		MINT 				&	2.1M		&	6M		& biopax-2:protein \\ \hline
25&		NCI-Nature 			&  	611k		&	237k	& biopax-2:protein \\ \hline
26&		Phenotype Ontology 	&	84k			& 	36k		& - \\ \hline
27&		Reactome 			& 	815k		&	330k	& biopax-2:protein \\ \hline
28&		Sider 				& 	102k		&	30k		& - \\ \hline
29&		Symptom				& 	4.2k		&	2k		& - \\ \hline
\end{tabular}
\caption{\footnotesize Lifescience datasets used for federation scenario: 29 datasets/$4$B+ RDF triples}
\label{tbl:datasets}
\end{table}

\textbf{Queries.}
We selected two query sets that implement realistic use cases on top of the life science data collection. 
The first query set (LS in Table~\ref{table:query_characteristics}) is a slightly modified version of the Life Science query set from the FedBench benchmark suite, updated to reflect changes in the schema and data of the latest versions of the respective data sets. The second query set (LLD in Table~\ref{table:query_characteristics}) contains sample queries from Linked Life Data (cf. \url{http://linkedlifedata.com/sparql}) and represents typical queries that can be performed against the integrated set of life science databases. We limited the selection to those queries that can be answered based on publicly available data sets  (i.e., without data exclusively available through the Linked Life Data system).  Figure~\ref{ref:tblan} exemplarily discusses two sample queries taken from the two query sets. 

Table~\ref{table:query_characteristics} gives an overview of the benchmark queries and their properties, showing that they vastly vary in their characteristics. In particular, we indicate the SPARQL operators that are used inside the query (\emph{Op.}), the solution modifiers that were used additionally (\emph{Sol.}), categorize the query structure (\emph{Struct.}), roughly distinguishing different join combinations -- like subject-subject or subject-object joins -- leading to different query structures commonly referred to as star-shaped, chain, or hybrid queries, and indicate the number of results (\emph{\#Res.}) on the federation datasets. 

\begin{table}[t]
\scriptsize
  \vspace*{-6ex}
  \caption{Summary of query characteristics. Operators: \textbf{A}nd (``.''), \textbf{U}nion, \textbf{F}ilter, \textbf{G}roup By, Count (\#), \textbf{O}ptional; Solution modifiers: \textbf{D}istinct, \textbf{L}imit, \textbf{Of}fset, \textbf{Or}der By; Structure: \textbf{S}tar, \textbf{C}hain, \textbf{H}ybrid}
\begin{center}
  \vspace*{-2ex}
  \hspace*{-4em}
  \begin{tabular}{|l|c|c|c|c|l|l|c|c|c|c|}
   \cline{1-5} \cline{7-11} 
    \multicolumn{5}{|c|}{\bf FedBench Life Science (LS)} & \ \ \ & \multicolumn{5}{|c|}{\bf Linked Life Data (LLD)} \\
   \cline{1-5} \cline{7-11} 
    & {\em Op.} & {\em Mod.} & {\em Struct.} & {\em \#Res.} &  &  & {\em Op.} & {\em Mod.} & {\em Struct.} & {\em \#Res.} \\   \cline{1-5} \cline{7-11} 
    {\bf 1} & 	U 	& - & - & 1159 	& 		& {\bf 1} & A 	& - & S	&  167 	\\   \cline{1-5} \cline{7-11} 
    {\bf 2} & 	AU	& - & - &  319 	&		& {\bf 2} & A 	& D & H	&  	22	\\   \cline{1-5} \cline{7-11} 
    {\bf 3} & 	A 	& - & H & 9869 	&		& {\bf 3} & A 	& - & C	& 	70	\\   \cline{1-5} \cline{7-11} 
    {\bf 4} & 	A 	& - & H & 3 	&		& {\bf 4} & A 	& D & C	&  210	\\   \cline{1-5} \cline{7-11} 
    {\bf 5} &	A 	& - & H & 395 	&		& {\bf 5} & A	& D & C	&  45	\\   \cline{1-5} \cline{7-11} 
    {\bf 6} & 	A 	& - & H & 28 	&		& {\bf 6} & AF	& D & H	&  63	\\   \cline{1-5} \cline{7-11}
    {\bf 7} & 	AFO	& - & H & 109 	&		& {\bf 7} & A	& - & H	&  59	\\   \cline{1-5} \cline{7-11} 
    \multicolumn{5}{l}{}					& 		& {\bf 8} & A 	& - & H	& 131	\\   \cline{7-11} 
	\multicolumn{5}{l}{}					& 		& {\bf 9} & \#	& - & -	&  	1	\\   \cline{7-11} 
    \multicolumn{5}{l}{}					& 		& {\bf 10}& AG 	& - & C	&  	2	\\   \cline{7-11} 
  \end{tabular}
  \vspace*{-4ex}
\end{center}
\label{table:query_characteristics}
\end{table}

\begin{figure}[t]
\scriptsize
\begin{center}
  \begin{tabular}{p{0.48\textwidth}p{1em}p{0.47\textwidth}}
  {\bf Example, Life Science Query 4:} For all drugs in DBpedia, find all drugs they interact with, along with an explanation of the interaction. & &
  {\bf Example, Linked Life Data Query 8:}  Select all  human genes located on the Y-chromosome with known molecular interactions. \\[-2.5ex]
\begin{verbatim}
SELECT ?Drug ?IntDrug ?IntEffect WHERE {
   ?Drug rdf:type dbpedia-owl:Drug .
   ?y owl:sameAs ?Drug .
   ?Int drugbank:interactionDrug1 ?y .
   ?Int drugbank:interactionDrug2 ?IntDrug .
   ?Int drugbank:text ?IntEffect . }
\end{verbatim} & & 
\begin{verbatim}
SELECT ?genedescription ?taxonomy ?interaction WHERE {
    ?interaction biopax2:PARTICIPANTS ?p .
    ?interaction biopax2:NAME ?interactionname .
    ?p biopax2:PHYSICAL-ENTITY ?protein .
    ?protein skos:exactMatch ?uniprotaccession .
    ?uniprotaccession core:organism ?taxonomy .
    ?taxonomy core:scientificName 'Homo sapiens' .
    ?geneid gene:uniprotAccession ?uniprotaccession .
    ?geneid gene:description ?genedescription .
    ?geneid gene:chromosome 'Y' . }
\end{verbatim} \\
 \\[2ex]
  \end{tabular}
 
\end{center}
\vspace*{-6ex}
\caption{Selected benchmark queries}
\vspace*{-3ex}
\label{ref:tblan}
\end{figure}

A complete description of the data sets (including download links) and queries used in the benchmark is available at \url{http://biofed.fluidops.net/}.

\section{Experiments}
\label{sec:experiments}

\subsection{Infrastructure Description and Setup}

In our experiments we focus on two different federated settings.
First, we set up a \textit{local federation} to evaluate the performance and practicability of federated data processing
with FedX in a controlled setting with low network latency, where all
endpoints are deployed in a dedicated local environment. Complementary, the \textit{hybrid federation} consists
of a mix of local and remote SPARQL endpoints (the latter hosted in the Amazon AWS cloud), which allows us to
study the implications arising in scenarios with higher network latency. The hybrid scenario reflects challenges in
the enterprise context, where private, enterprise-internal data sources are combined with public SPARQL endpoints
in a federated setting.
To guarantee repeatability of the experiments we establish a controlled environment in both settings,
i.e. we use SPARQL endpoints running on non-shared compute and storage resources. The details are descibed
in the following.

{\bf Local federation.}
For the local federation we provide access to the lifescience datasets through individual SPARQL endpoints running
in our local computing cluster. In this cluster we use two HP Proliant DL360 servers running a 64bit Windows Server
operating system, one with 8x2GHz CPU and 64GB RAM (\texttt{Server1}), the other with 2x3GHz CPU and 20GB RAM
(\texttt{Server2}), both backed by fast storage. The total available memory is distributed to the individual SPARQL
endpoints corresponding to the number of triples, e.g. the Uniprot endpoint got assigned a total memory of 14GB,
while the smaller Drugbank endpoint is running in a 1.5GB process. The datasets 1 to 14 from \tbl{tbl:datasets} are
deployed on \texttt{Server1} and the remaining ones, 15 to 29, are deployed on \texttt{Server2}.

The individual SPARQL endpoints are powered by a state-of-the-art triple store implementing the OpenRDF Sesame
interface\footnote{http://www.openrdf.org},  running in Tomcat 6 application server processes. Sesame is the
de-facto standard framework for processing RDF data and offers access to RDF storage solutions through
an easy-to-use API. The triple stores themselves can be accessed via  SPARQL endpoints.

{\bf Hybrid federation.}
For the hybrid setting we deployed selected SPARQL endpoints from the local infrastructure to an Amazon AWS
EC2 instance. More precisely, we deployed the DrugBank, Uniprot, and Pubmed data sets in the AWS cloud.
Like in the local setting, these data sets were deployed as individual SPARQL endpoints on top of a
Tomcat 6 application server, using exactly the same
database setup and memory assignment for the individual endpoints as in the local setting. The endpoints
were hosted together on a single, high-memory AWS instance (type ``m2.2xlarge'') running 64bit MS Windows Server 2008
with 13 EC2 Compute Units (4 virtual cores with 3.25 EC2 compute units each) and 34.2GB memory.
The data sets were attached to the instance using Amazon EBS storage volumes. The instance and
volumes were both hosted in the AWS zone ''EU West (Ireland)'', allowing for fast communication between compute
and storage infrastructure. Note that Ireland is the AWS zone closest to Germany, where the local
endpoints and FedX were run.

{\bf Mediator and benchmark driver.}
In both settings, the federation was driven by the FedX v2.0 federation mediator described in Section~\ref{sec:technology}.
FedX was configured to run over the set of the $29$ Bio2RDF SPARQL endpoints, 
either using only local endpoints (in the local setting) or the combination of local and global endpoints
described above in the federated setting. For running the experiments we used
FedBench\footnote{FedBench project page: http://code.google.com/p/fbench/} \cite{schmidt2011a}, a comprehensive benchmark suite for
analyzing the efficiency and effectiveness of federated query processing strategies over semantic data that provides
customizable benchmark drivers.

{\bf Metrics.}
The central measure in our experiments is the query evaluation time: in both the local and the hybrid
scenario we report on the average elapsed time over five runs, assessed after five previous warmup runs.
Following the guidelines described in~\cite{DBLP:journals/cacm/FlemingW86}, we indicate the geometric mean, which is defined
as the n$^{th}$ root over the product of $n$ values: compared to the arithmetic mean, the geometric mean
flattens outliers, which -- in our setting --  occasionally arised, particularly in the hybrid federation, due to
unpredictable effects such as punctually high network delays. Other metrics we discuss are (i)~the number
of requests sent to SPARQL endpoints from FedX during query evaluation, (ii)~the number of triple patterns in
the individual queries and (iii)~the efficiency of the source selection algorithm in FedX.
As we will discuss in the following, these are parameters that
have significant influence on the query evaluation times.

\subsection{Experimental Results}
\label{subsec:results}

\begin{figure}[t]
\includegraphics[width=0.9\textwidth]{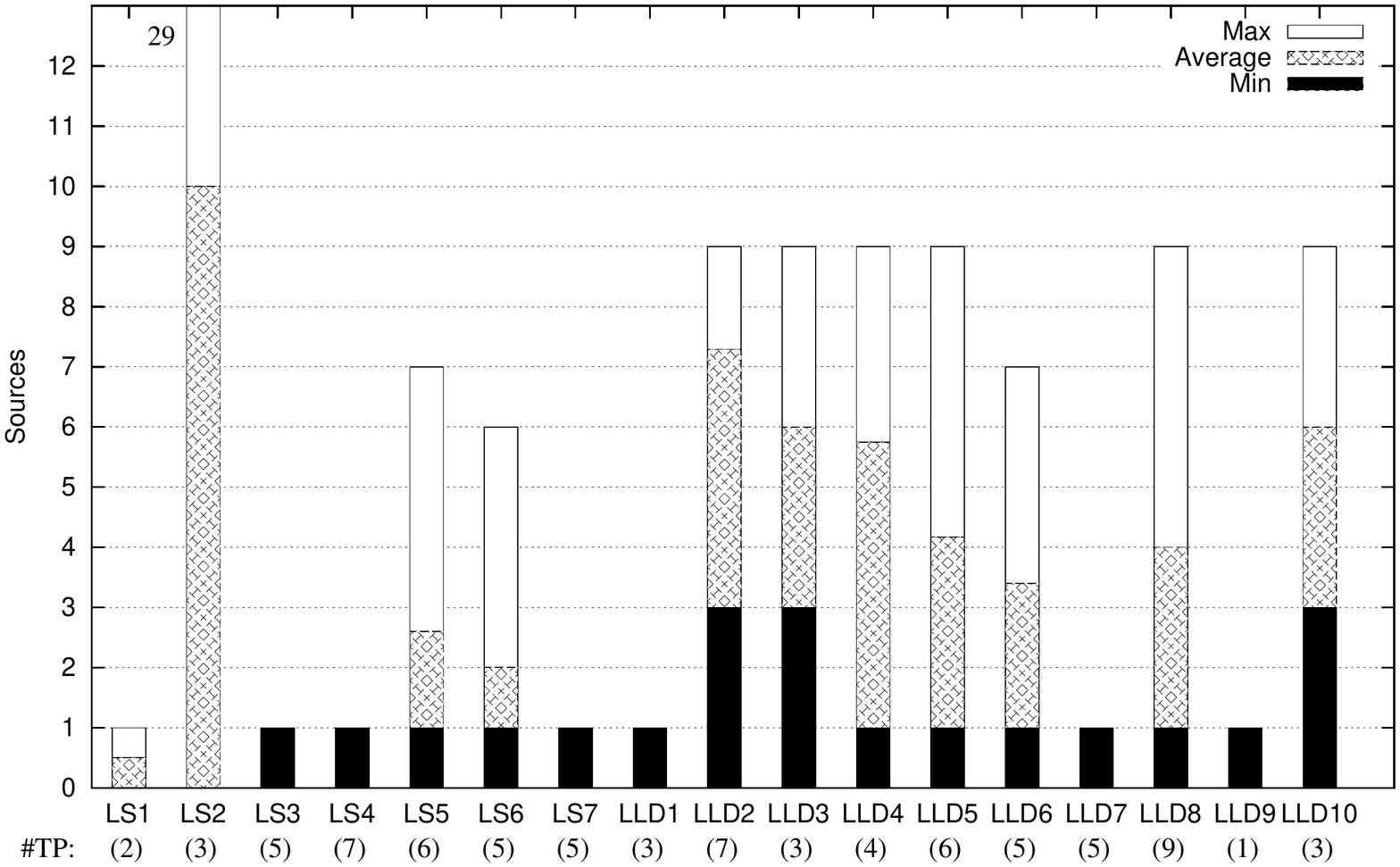}
\vspace{-0.4cm}
\caption{Source selection analysis: relationship between queries and triple patterns w.r.t.~relevant endpoints according to FedX' source selection algorithm.}
\label{fig:source_selection}
\end{figure}

We start with a discussion of FedX' source selection strategy (cf.~Section~\ref{sec:technology}), which
forms the basis for the understanding of the subsequent results. In order to minimize
the number of requests, FedX -- prior to evaluating the query -- sends \textsc{ASK} queries for the triple patterns
contained in the query to all SPARQL endpoints, to identify which sources are potentially relevant for
which patterns in the query. This information is then used to optimize query processing, such as 
sending patterns only to relevant endpoints or grouping subqueries that can be answered by a single
endpoint alone. Visualizing the outcome of the source selection strategy, Figure~\ref{fig:source_selection} shows,
for each of the benchmark queries (i)~the number of triple patterns in the query (plotted below the query name)
and (ii)~the minimum, maximum, and average (over all triple patterns in the query) number of endpoints
that have been identified as relevant for the triple patterns according to the source selection strategy.
As an example, query LLD2 is composed out of $7$ triple patterns, where the minimal pattern(s) retrieve
non-empty results from $3$ endpoints, the maximal pattern(s) retrieve results from $9$ endpoints,
and the average number of endpoints that contribute results to a triple pattern in LLD2 is about $7.2$.
As a whole, the diagram leads to two interesting observations: first of all, the queries vary in complexity
regarding the number of sources that (potentially) contribute to the query result: there are simple queries
which can be answered by querying a single source, while others have triple patterns containing potential
matches in up to $9$, in the worst case even all $29$ federation members\footnote{Query $LS2$ contains
the triple pattern \textsc{?caff ?predicate ?object}, which -- taken alone -- can be answered by all endpoints.}.
Second, the results demonstrate that the source selection strategy of FedX is quite efficient, reducing the
average number of sources involved in answering triple patterns to at most $10$ out of $29$ for all the
queries, typically even less. Given that the number of requests is one of the main factors driving evaluation time
(as will be discussed in the following), this efficient source selection strategy can be seen as a cornerstone for
 the practicability of FedX.

\begin{figure}[h!]
 \scriptsize
 \begin{minipage}{1\textwidth}
 \textbf{(a)} \\[-2cm]
 \includegraphics[width=1.1\textwidth]{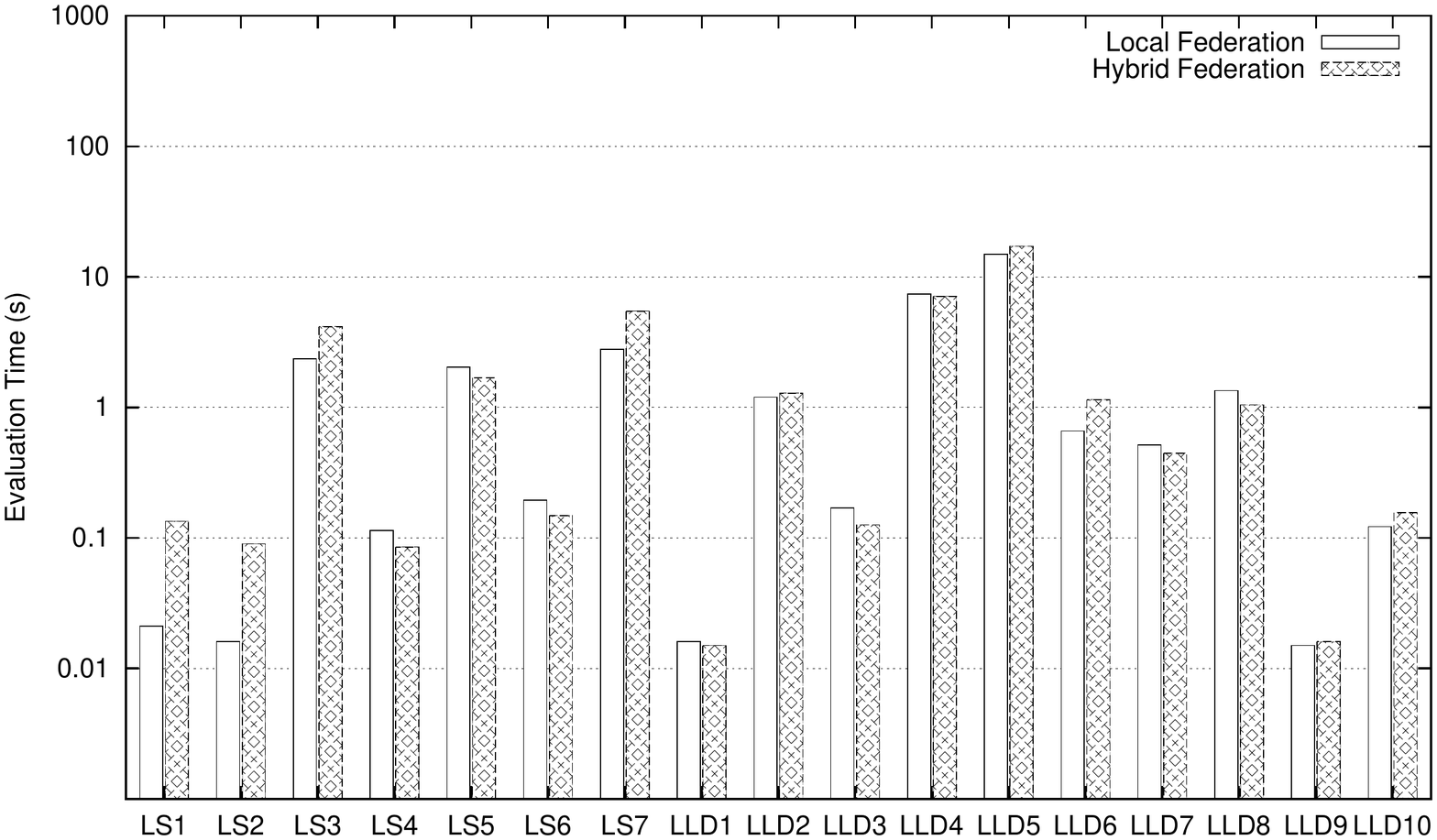}
 \end{minipage} 
 \begin{minipage}{0.54\textwidth}
 
 \vspace*{-1.2cm}
 \begin{tabular}{|l|c|c|c|c|c|}
 \multicolumn{6}{l}{\textbf{b)}} \\
 \hline
 			&	\textbf{Local}		& \textbf{Hybrid}  	&  \textbf{\#Req} 		& \textbf{\#Req(D)}		& \textbf{\#Req(U)}	\\ \hline
LS1		 	&	0.021			& 0.134			&	1				& 1					& 0				\\ \hline
LS2		 	&	0.016			& 0.090			&	1				& 1					& 0				\\ \hline
LS3			&	2.356			& 4.159			&	1512			& 1511				& 0				\\ \hline
LS4		 	&	0.114			& 0.085			&	3				& 1					& 0				\\ \hline
LS5		 	&	2.037			& 1.678			&	815				& 1					& 91				\\ \hline
LS6		 	&	0.194			& 0.148			&	84				& 1					& 0				\\ \hline
LS7		 	&	2.783			& 5.451			&	1355			& 110				& 0				\\ \hline 
LLD1 		&	0.016			& 0.015			&	1				& 0					& 0				\\ \hline 
LLD2 		&	1.199			& 1.282			&	649				& 0					& 0				\\ \hline 
LLD3 		&	0.170			& 0.126			&	75				& 0					& 0				\\ \hline 
LLD4		&	7.358			& 7.067			&	3043			& 77					& 78				\\ \hline 
LLD5	 	&	14.823			& 17.167			&	6301			& 806				& 109			\\ \hline 
LLD6	 	&	0.660			& 1.142			&	135				& 21					& 19				\\ \hline 
LLD7	 	&	0.514			& 0.446			&	162				& 0					& 161			\\ \hline 
LLD8		&	1.345			& 1.045			&	521				& 9					& 148			\\ \hline 
LLD9 		&	0.015			& 0.016			&	1				& 0					& 0				\\ \hline 
LLD10 		&	0.122			& 0.156			&	75				& 0					& 0				\\ \hline 
 \end{tabular}
 \end{minipage}
\hspace{0.15cm}
  \begin{minipage}{0.45\textwidth}
   \vspace*{-1.2cm}
 \begin{tabular}{|l|c|c|c|}
 \multicolumn{4}{l}{\textbf{c)}} \\
 \hline
 	&	\textbf{No Caching}	& \textbf{Caching}  &  \textbf{\#Savings} 	\\ \hline
LS1 &	0.203			& 0.134			&	58				\\ \hline
LS2 &	0.309			& 0.090			&	87				\\ \hline
LS3 &	4.274			& 4.159			&	145				\\ \hline
LS4 &	0.461			& 0.085			&	203				\\ \hline
LS5 &	2.098			& 1.678			&	174				\\ \hline
LS6 &	0.462			& 0.148			&	145				\\ \hline
LS7 &	5.434			& 5.451			&	145				\\ \hline 
LLD1 &	0.440			& 0.015			&	87				\\ \hline 
LLD2 &	2.162			& 1.282			&	203				\\ \hline 
LLD3 &	0.429			& 0.126			&	87				\\ \hline 
LLD4 &	7.077			& 7.067			&	116				\\ \hline 
LLD5 &	16.952			& 17.167			&	174				\\ \hline 
LLD6 &	1,098			& 1.142			&	145				\\ \hline 
LLD7 &	0.066			& 0.446			&	144				\\ \hline 
LLD8 &	1.511			& 1.045			&	261				\\ \hline 
LLD9 &	0.063			& 0.016			&	29				\\ \hline 
LLD10 &	0.311			& 0.156			&	87				\\ \hline 
 \end{tabular}
 \end{minipage}
\caption{Experimental Results: (a)~Graphical comparison of query evaluation time in local and hybrid federation; (b)~Tabular listing of evaluation times and number of requests sent to SPARQL endpoints during query evaluation; (c)~Influence of source selection caching in FedX on evaluation times in the hybrid setting.}
\label{fig:experiments}
\end{figure}

Figure~\ref{fig:experiments}(a) compares the query evaluation times for our $17$ benchmark queries over the local and hybrid
federation, with source selection caching enabled. Given the warmup phase prior to taking the measurements,
an active source selection cache implies that FedX in this scenario has full knowledge about which
sources can contribute results to which triple pattern in the input query. Starting with the discussion of the local federation setting,
we can observe that all $17$ queries return a result within $15$s. $15$ queries are faster than $3s$,
$10$ queries in the subsecond range, and $5$ queries are even faster than $0.1$s. Given that our queries represent 
a mix of dedicated benchmark queries designed particularly to test challenges in federated scenarios and real-world use
cases from the Bio2RDF project, these numbers impressively demonstrate the practicability of FedX as a mediator
for large-scale RDF federations in the billion triple range.

In addition to a tabular representation of the evaluation time for the two settings (columns \textbf{Local} and
\textbf{Hybrid}), Figure~\ref{fig:experiments}(b) summarizes the number of requests (\textbf{\#Req}) sent to
the SPARQL endpoints during query evaluation. We can observe a clear coincidence between the number of requests
sent during query evaluation and the query evaluation time; for instance, the five most expensive queries (in terms
of runtime) -- LS3, LS5, LS7, LLD4, LLD5 -- are characterized by the five highest numbers of requests sent to
SPARQL endpoints. This indicates that the network delay is the dominating factor in query evaluation.

In order to study the effect of network latency in more detail, we next compare the results in the local federation
with the hybrid federation. As expected, the query results in the hybrid setting are generally (yet not always)
slower due to the higher network latency induced by the communication with the SPARQL endpoints in the AWS cloud.
Going into more detail, Figure~\ref{fig:experiments}(b) also shows
the number of subqueries sent against the remote SPARQL endpoints Drugbank (\textbf{\#Req(D)}) and Uniprot
(\textbf{\#Req(U)}).
Based on these numbers, we can classify the queries into three classes. The first class contains queries that do not
require communication between FedX and the remote endpoints (LLD1-3 and LLD9-10). For these queries, FedX' source selection
cache helps to avoid expensive requests to the endpoints in the AWS cloud, so we observe no or only small
overheads in evaluation time. The second class of queries, such as LS3, LS7, and LLD5, require a considerable amount
of requests against the remote endpoints; as a consequence, the execution time in the federated setting increases. Still, the
highest percental increase of about 100\% can be observed for LS7, which still results in practical response times.
Somewhat surpisingly, we can observe a third class of of queries, for which the hybrid setup even outperforms the
local setup (LS4-6, LLD1, LLD3-4, and LLD7-8). This result can be explained by the fact that the overall load on the
Amazon machine, which hosts only three endpoints -- rather than $14-15$ endpoints, as it is the case for the two local servers --
is lower, which in turn results in generally faster response times for the subqueries sent to the endpoints. This shows that in many cases
the advantages gained by distribution dominate the overhead imposed by increased communication costs.

Finally, in Figure~\ref{fig:experiments}(c), we study the influence of source selection caching in FedX:
the first two columns compare the evaluation times of the queries with source selection caching disabled vs.~enabled
in the hybrid setting; the \textbf{\#Savings} column denotes the number of \textsc{Ask} requests
(sent to endpoints in order to find out whether they can contribute answers to a given triple pattern) that could be
saved when caching was turned on. As can be seen, caching leads to runtime savings in most cases. As a general
trend, the percentual savings are particularly high whenever \textbf{\#Savings} is high compared to the overall
number of requests, \textbf{\#Req}, depicted in Figure~\ref{fig:experiments}(b); for instance, for query LS2, we
save $87$ requests (out of, in total, $87$+$1$ = $88$ requests to endpoints), which induces a significant percental speedup.
For queries where the number of requests is already high (e.g., LS3, LS7, or LLD4), the caching benefits are negligible.
In summary, the results show that the source selection cache is not crucial for efficient evaluation, thus proving the
flexibility of FedX which allows to add new federation members ad hoc, without warming up caches or precalculating
statistics. Still, source selection caching yields an additional speedup for most queries, which can be particularly
beneficial in scenarios with high query loads involving many simple queries.

\section{Conclusions and Future Work}
\label{sec:conclusions}

We presented the first large-scale RDF federation in the billion triple range over Bio2RDF data sources,
driven by the highly optimized Linked Data federation mediator FedX. Our exhaustive and repeatable
experimental evaluation demonstrates the practicability of our approach and studies various aspects
driving evaluation time. One cornerstone of evaluation performance is an efficient source selection
strategy. It is crucial to minimize the number of requests sent to the individual SPARQL endpoint
during query evaluation, which is the major bottleneck in efficient federated query processing.
Going beyond this finding, our experiments identify settings in which the advantages gained by
distribution dominate the overhead imposed  by increased communication costs, thus leveraging the
benefits of a federated setup with autonomous compute endpoints.

The focus of future work in this area  therefore  should lie on techniques to further minimize the
communication efforts. One promising approach, which we identified during our interpretation of
query evaluation plans, is to exploit data set specific  namespaces in URIs, in order to further improve
the source selection process. Another promising approach aiming at a combination of the benefits of
federation and centralization would be the automated colocation of data sets that exhibit frequent joins
and therefore impose high communication costs, which could e.g.~be reached by an adaptive query log
analysis, combined with a caching layer maintained inside the federation layer.

\bibliographystyle{plain}
\bibliography{main}

\end{document}